\documentstyle[aps,multicol,epsf,psfig]{revtex}
\newcommand{\ket}[1]{| #1 \rangle}
\begin{document}

\title{How many photons are needed to distinguish two transparencies?}
\author{ Graeme Mitchison$^{1}$, Serge Massar$^{2}$ and Stefano Pironio$^{2}$}
\address{$^{1}$MRC Laboratory of Molecular Biology, Hills Road, Cambridge CB2
2QH, UK.\\ $^{2}$Service de Physique Th\'eorique,
Universit\'e Libre de Bruxelles, CP 225, Bvd. du Triomphe, B1050
Bruxelles, Belgium.\\}
\date{Preliminary version -- \today}

\maketitle

\begin{abstract}
We give a bound on the minimum number of photons that must be absorbed by any
quantum protocol to distinguish between two transparencies. We show how a quantum Zeno method in
which the angle of rotation is varied at each iteration can attain this bound
in certain situations.
\end{abstract}

\vspace{0.2cm}

\begin{multicols}{2}

Making images of objects plays an important part in present day
science and technology. In certain situations the object may be
damaged by the radiation used to make the image. This can happen, for
example, in various types of microscopy of biological specimens.

The simplest way to make an image is to send light through the object
and measure how much is absorbed and how much is transmitted. However
by using the quantum properties of light and in particular using
interferometric techniques one can hope to decrease the amount of
radiation absorbed by the object. This question has attracted
considerable attention recently \cite{EV,Kwiat,Jang,KSS,1,MMP,KW,FHKPR}.

In this paper we shall consider the particular problem of distinguishing two
transparencies. This problem is sufficiently simple that it allows detailed
analytical treatment but is also sufficiently general that it gives a clear
insight into the advantages quantum interferometric techniques can bring to
minimal absorption measurements.

Thus suppose there are two objects, with amplitudes for transmission
of light $\alpha_1$ and $\alpha_2$. For instance $\alpha_1$
($\alpha_2$) could correspond to the presence (absence) of features
with distinctive density in a microscope preparation. Furthermore we
shall suppose that there are known prior probabilities $p_1$ and $p_2$
for objects 1 and 2, respectively. Then it will be shown that
the mean number of absorbed photons $\bar N^{abs}$ needed by any
quantum protocol to distinguish the two objects must satisfy the
following bound:
\begin{equation}
\label{result} {\bar N^{abs}} \ge {2|\beta_1\beta_2|
\left ( \sqrt{p_1p_2}-\sqrt{P_E(1-P_E)} \right ) \over (|1-\bar\alpha_1
\alpha_2|-|\beta_1\beta_2|)},
\end{equation}
where $\beta_i$ is the amplitude for absorption by object $i$ (so
$|\alpha_i|^2+|\beta_i|^2=1$ for $i=1, 2$), and $P_E$ is the probability of
error. For instance, this tells us that, with equal prior probabilities and
$P_E=0$, at least 175 photons must be absorbed in order to distinguish
$\alpha_1=0.2$ and $\alpha_2=0.3$, and at least 2.3 photons to distinguish
$\alpha_1=0.2$ and $\alpha_2=0.8$.

We have also shown, using numerical analyis, that this bound can be approached
arbitrarily closely in the case where $\alpha_1$ and $\alpha_2$ are real, and
when the prior probabilties are equal ($p_1=p_2=1/2$).

When the two transparencies are very close, the dependance of our
bound on $\alpha_1$ and $\alpha_2$, for fixed $P_E$, is similar
to that derived from classical counting of absorbed photons or simple
interferometric techniques \cite{MMP}. But quantum algorithms have the
noteworthy feature that they allow zero error probability. When one of
the $\alpha_i$ is zero, our bound is zero, and one is in the domain of
``interaction-free'' measurement \cite{EV}, where the probability of
photon absorption can be made arbitrarily small
\cite{Zeno,Kwiat,Jang,1}. Our new bound is interesting in that it
spans the entire range from almost identical transparencies to
``interaction-free'' measurement.

We now turn to the proof of our bound.  We follow \cite{MMP} and write
the Hilbert space of a general quantum protocol as a product of three
subspaces $H_{A}\otimes H_{P}\otimes H_{O}$. $H_{A}$ is the space of
ancillary photons which do not interact with the object and
$H_{P}$ the Fock space of the interrogating photons which are directed
through the object (for instance, $H_{A}$ corresponds to the empty arm
and $H_{P}$ to the object arm in the usual ``interaction-free''
measurement scheme). $H_{O}$ is the space of the object, with states
$\ket {n_{1},\ldots ,n_{j},\ldots}_{O}$ if $n_{1},\ldots n_{j},\ldots$
photons have been absorbed by the object at stages $1, \ldots, j,
\dots$ of the protocol. If object $i$ is present, the state at step
$j$ of the protocol can be written
\begin{equation}
\ket {\Psi
^{j}_{i}}=\sum _{k,m,\bf n}C^j_{i,km \bf n}\ket {k}_{_{A}}\ket {m}_{_{P}}\ket
{n_{1},\ldots n_{j-1},0_{j},0_{j+1},\ldots }_{_{O}},
\label{state}
\end{equation}
where $\ket {k}_{_{A}}$ denotes $k$ ancillary photons, $\ket {m}_{_{P}}$, $m$
interrogating photons and where the sum over ${n_1,\ldots,n_j-1}$,
$\sum_{n_1,\ldots,n_j-1}$, has been shortened to $\sum_{\bf n}$. The protocol is
assumed to consist of a sequence of unitary steps acting on the
joint subspace $H_{A}\otimes H_{P}$, alternating with steps where the
interrogating photons interact with the object. Finally, a measurement is
carried out whose result indicates which object is present.

The absolute value of the overlap $f^j$,
\begin{equation}
f^j=|\langle \Psi ^{j}_{1}|\Psi
^{j}_{2} \rangle|=|\sum _{k,m,\bf n}(\overline{C}_1^j C^j_2)_{mk\bf n}|.
\end{equation}
between the states for objects 1 and 2 measures how effectively the protocol
can distinguish the two objects at step $j$. The overlap $f^j$ is not altered
by unitary steps, of course, but is by interaction steps.
The interaction step for object $i$ can be described by
$a^{\dagger }_P\rightarrow \alpha_i a^{\dagger }_P+\beta_i
b^{\dagger }_O$, where $a^{\dagger }_P$ and $b^{\dagger }_O$ are the
creation operators in $H_P$ and $H_O$, respectively. Then one can show
(see \cite{MMP} for details) that after the interaction step
\begin{equation}
\label{prodsc}
f^{j+1}=|\sum _{k,m,\bf n}(\overline{C}_1^j C^j_2)_{mk\bf n}(\bar\alpha_1
\alpha_2+\bar\beta_1\beta_2)^m|.
\end{equation}
The idea of the proof is to put a bound on the difference $\Delta
f=f^{j}-f^{j+1}$. Since
\begin{eqnarray}
\lefteqn{\Delta f=f^{j}-f^{j+1} } \nonumber\\
& =&|\sum _{k,m,\bf n}(\overline{C}_1^j C^j_2)_{mk\bf
n}|-|\sum _{k,m,\bf n}(\overline{C}_1^j C^j_2)_{mk\bf n}(\bar\alpha_1
\alpha_2+\bar\beta_1\beta_2)^m|
\nonumber \\
&\le &|\sum _{k,m,\bf n}(\overline
C^j_1C^j_2)_{km \bf n}\{1-(\bar\alpha_1
\alpha_2+\beta_1\beta_2|)^m\} \label{firststep2}
\\
& \le & \sum_{k,m,\bf n}|(\overline C^j_1C^j_2)_{km\bf n}||\{1-(\bar\alpha_1
\alpha_2+\beta_1\beta_2)^m\}|,
\label{firststep3}
\end{eqnarray}
it is a useful intermediate step to find a bound for $|1-(\bar\alpha_1
\alpha_2+\bar\beta_1\beta_2)^m|$. Now the phases of the $\beta$
coefficients are inaccessible to experiments since they are the phases
accumulated by the macroscopic object when it absorbs a photon. Hence
we can choose the phase of $\beta_i$ that gives the tightest
bound. This is the motivation for the following:\\

There is a value of $\phi$ such that, for all
integers $m \ge 1$,
\begin{equation}
|1-(\bar\alpha_1
\alpha_2+e^{i\phi}|\beta_1\beta_2|)^m| \le m(|1-\bar\alpha_1 \alpha_2|-|\beta_1\beta_2|).
\label{del}
\end{equation}

\noindent {\em Proof } Putting $\sigma=(\bar\alpha_1
\alpha_2+e^{i\phi}|\beta_1\beta_2|)$, it is easy to check that
$|1-\sigma|$ is minimized by taking
\[
(1-Re(\bar\alpha_1\alpha_2))\sin \phi =-Im(\bar\alpha_1\alpha_2)\cos \phi ,
\]
with $-{\pi \over 2} \le \phi \le {\pi \over 2}$, and, for this value of
$\phi$, $|1-\sigma|=|1-\bar\alpha_1 \alpha_2|-|\beta_1\beta_2|$. This
establishes the hypothesis for $m=1$. Assume (\ref{del}) holds for
$m$. Then, with the same value of $\phi$,
\begin{eqnarray*}
|1-\sigma^{m+1}| &=&  |1-\sigma + \sigma (1-\sigma^m)|\\
&\le& |1-\sigma| + |\sigma| |1-\sigma^m|\\
 &\le& |1-\sigma|+m|\sigma||1-\sigma| \\
&=& (1+m|\sigma|)|1-\sigma|\\
&\le&(m+1)|1-\sigma|,
\end{eqnarray*}
where we have used $|\sigma|\le 1$. This
establishes the hypothesis for all $m$ and proves inequality (\ref{del}).

Suppose that we have chosen the phases of the $\beta_i$ so that
$\beta_1\beta_2=e^{i\phi}|\beta_1\beta_2|$, where $\phi$ is chosen so
that (\ref{del}) holds. We can rewrite (\ref{firststep3}) as
\begin{eqnarray}
\Delta f & \le & \sum_{k,m,\bf n}|(\overline C^j_1C^j_2)_{km\bf
n}||1-(\bar\alpha_1 \alpha_2+e^{i\phi}|\beta_1\beta_2|)^m|
\\
\label{firststep4}
& \le& |1-\bar\alpha_1 \alpha_2|-|\beta_1\beta_2|)\sum _{k,m,\bf
n}|(\overline C^j_1C^j_2)_{km\bf n}|m,
\end{eqnarray}
Writing $\gamma=(|1-\bar\alpha_1 \alpha_2|-|\beta_1\beta_2|)/
|\beta_1\beta_2|$, we have
\begin{eqnarray}
\Delta f & \le& \gamma \sum _{k,m,\bf n}|(\overline {\beta_1
C^j_{1,km\bf n}}\beta_2C^j_{2,km\bf n})|m
\nonumber \\
\label{secondstep2}
& \le&{\gamma \over 2\sqrt{p_1p_2}}
\sum _{k,m,\bf n} (p_1|\beta_1 C^j_{1,km\bf n}|^2+p_2|\beta_2 C^j_{2,km\bf
n}|^2)m
\\ \nonumber
\end{eqnarray}
since $|xy|\leq {p_1|x|^2+p_2|y|^2 \over 2\sqrt{p_1p_2}}$.
The last equation can be rewritten as
\begin{eqnarray}
\label{secondstep3}
\Delta f &\le &{\gamma \over
2\sqrt{p_1p_2}}(p_1n^{abs,j}_1+p_2n^{abs,j}_2),
\end{eqnarray}
where $n^{abs,j}_i$ is the expected number of photons absorbed at
step $j$ with object $i$ present. Starting from $f^1=1$ and iterating
gives
\begin{eqnarray}
1-f^K &\le&{\gamma \over 2 \sqrt{p_1p_2}} \sum_{j=1}^{K-1}
\left(p_1n^{abs,j}_1+p_2n^{abs,j}_2 \right)
\nonumber \\
&=&1- {\gamma \over 2
\sqrt{p_1p_2}}\left({p_1\bar N^{abs}_1+p_2\bar N^{abs}_2}\right),
\nonumber \\ \nonumber
\end{eqnarray}
where $N^{abs}_i$ is the expected number of photons absorbed when
object $i$ is present. Inserting the value of $\gamma$ and using
$\bar N^{abs}=p_1\bar N^{abs}_1+p_2\bar N^{abs}_2$ we get
\[
{\bar N^{abs}} \ge {2|\beta_1\beta_2| \sqrt{p_1p_2}(1-f^K)\over
(|1-\bar\alpha_1 \alpha_2|-|\beta_1\beta_2|)}.
\] Substituting $f^K=\sqrt{{P_E(1-P_E)}\over p_1p_2}$ (see \cite{Hels},
chapter IV 2, and in particular eq 2.34), completes the proof of
inequality (\ref{result}).

\begin{figure}[h]
\centerline {\psfig{width=9.0cm,file=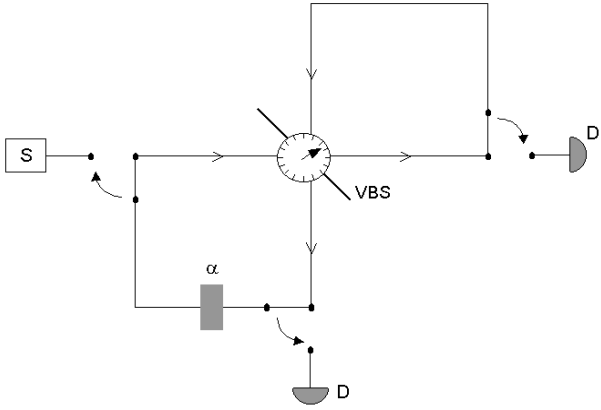}}
\begin{narrowtext}
\caption{A setup that allows two transparencies to be distinguished
with an arbitrarily close approximation to the minimum number of
absorbed photons, in the case where the two transparencies
$\alpha_{1,2}$ are real and have equal a priori probabilities. A
single photon source {\bf S} sends {\em one} photon into the
apparatus. After the photon has entered, it cycles through the
apparatus a certain number of times. In the apparatus the photon first
encounters a variable beam splitter (VBS) which sends the photon along
two different paths.  The object (shaded rectangle), with
transparency $\alpha=\alpha_1$ or $\alpha_2$, is inserted in
one of the paths. The initial reflectivity of the VBS is arbitrary but
taken to be very small.  Thereafter the reflectivity of the VBS
changes as described in the text.  After having cycled through the
apparatus a certain number of times, the photon is sent to one
of two detectors depending on which path the photon takes. Note that
the photon first passes through the VBS before being sent to the
detectors. This ensures that the measurement basis coincides with the
path the photon is on. If the detectors do not detect any photon, the
experiment is run again until one of the detectors registers a
photon. The variable beam splitter and switches used in the setup can
be implemented for instance by a Mach-Zehnder interferometer with a
variable phase-shifter placed in one of its arms.
\label{F1}} \end{narrowtext} \end{figure}

Is our bound optimal? In other words, is there a protocol that attains
the limit imposed by (\ref{result})? For real $\alpha_i$ and with
equal prior probabilities ($p_1=p_2=1/2$), we show this is the case
(more precisely, that the bound can be approached arbitrarily
closely). The protocol we use is a single photon protocol based on the
quantum Zeno ``interaction-free'' measurement scheme but where the
angle of rotation is now \emph{varied} at each iteration (see
figure \ref{F1}).  The photon is initially fed into a Mach-Zehnder
interferometer with the object placed in one of its arm. The photon
traverses the interferometer $K$ times, and is finally detected by two
detectors placed at each output port, provided it has not been
absorbed. If the photon is absorbed, the protocol is repeated until a
measurement outcome is obtained. As we will see below, no information
about the transparency is obtained from the absorption of a photon
because the probability of absorbing a photon at each step is the same
for both objects.

Let us analyse the protocol with the formulation used in the proof of
the bound.  At step $j$ of the protocol, we are in the state \[
\ket{\Psi^j_i}=a^j_i \ket{1}_A \ket{0}_P
\ket{0}_O+b^j_i\ket{0}_A\ket{1}_P\ket{0}_O+\ket{I^j}\ ,
\]
where $\ket{1}_A \ket{0}_P \ket{0}_O$ corresponds to the photon
being present in the empty arm and $\ket{0}_A\ket{1}_P\ket{0}_O$ to
the photon being in the object arm, and where $\ket{I^j}$ indicates
terms where interaction with the object has occurred on previous
steps. After the interaction step and the unitary step which acts on
$H_A \otimes H_P$ by the rotation $\left(\matrix{\cos \theta^j & -\sin
\theta^j \cr \sin \theta^j & \cos \theta^j \cr}\right)$, the state
becomes
\begin{eqnarray}
\label{nextstate}
\ket{\Psi_i^{j+1}}&=&(a^j_i \cos
\theta^j -\alpha_ib^j_i \sin \theta^j)\ket{1}_A \ket{0}_P \ket{0}_O \nonumber \\
& & \mbox{} +(\alpha_ib^j_i \cos \theta^j + a^j_i \sin \theta^j)
\ket{0}_A\ket{1}_P\ket{0}_O\nonumber \\
& & \mbox{} +\beta_i b^j_i \ket{0}_A\ket{0}_P\ket{1}_O\}+ I^j,
\end{eqnarray}

In order to attain our bound we shall examine the different
inequalities occurring between steps (\ref{firststep2})
and (\ref{secondstep3}) and try to saturate them. 
Note that because there is only one photon present, there are several
simplifications. 
There is only one term under the sum in
(\ref{firststep2}), and thus (\ref{firststep3}) is an
equality. Furthermore, $m=1$ in this term, and for $m=1$ (\ref{del})
is an equality. So (\ref{firststep4}) is also an equality. There are
two remaining places where an inequality could occur, namely
(\ref{firststep2}) and (\ref{secondstep2}). The first of these will be
an equality if
\begin{eqnarray}
& & \langle \Psi ^{j}_{1}|\Psi^{j}_{2} \rangle, \langle \Psi
^{j+1}_{1}|\Psi ^{j+1}_{2} \rangle
\nonumber \\
& & \mbox{ and } \langle \Psi ^{j}_{1}|\Psi
^{j}_{2} \rangle-\langle \Psi ^{j+1}_{1}\Psi
^{j+1}_{2} \rangle
\nonumber \\
& & \mbox{ are simultaneously all}  \ge 0  \mbox{ or }
\le 0,
\label{triangle}
\end{eqnarray}
and the second if $|\beta_1C^j_{1,km \bf n}|=|\beta_2 C^j_{2,km\bf
n}|$, (we are assuming $p_1=p_2$). For real $\alpha_i$, the
$a^j_i$ and $b^j_i$ will all be real, and the latter condition amounts
to
\begin{eqnarray}
\label{condition}
\beta_1(\alpha_1b^j_1 \cos \theta^j + a^j_1 \sin
\theta^j)&=&\beta_2(\alpha_2b^j_2 \cos \theta^j + a^j_2 \sin \theta^j),
\end{eqnarray}
or
\begin{equation}
\label{variable}
\tan \theta^j={{\alpha_1\beta_1 b^j_1-\alpha_2\beta_2 b^j_2} \over {\beta_2
a^j_2-\beta_1 a^j_1}}.
\label{tj}
\end{equation}

A protocol starts with the initial state $\ket{1}_A \ket{0}_P \ket{0}_O$
so that $b_1^0=b_2^0=0$. Thus the angle given by (\ref{variable}) is
$\theta^0=0$, so no photon ever passes through the object. To avoid this, a
first angle of rotation $\theta^0\ne$ 0 must be chosen. At the first step,
therefore, equality in (\ref{secondstep2}) cannot be achieved. For subsequent
steps, however, rotations according to (\ref{variable}) are applied. If
$\theta^0$ is small, the departure from equality in
(\ref{secondstep2}) will be small. However, we can only expect a near approach
to our bound, not equality. Note that the condition (\ref{condition}) just says
that the probabilities $|\beta_ib_i^{j+1}|^2$ of absorbing a photon are the
\emph{same} for both objects. This means that no
information about one of the objects is obtained by the absorption of a photon.

We ran a computational test on our variable angle algorithm
(\ref{variable}) to explore its behaviour. The two parameters to
choose are the initial angle $\theta^0$ and the number of iteration
steps $K$. $\theta^0$ was chosen randomly (and was taken very
small). The maximum number of steps allowed is determined by the
condition (\ref{triangle}). Since initially $\langle \Psi
^{0}_{1}|\Psi^{0}_{2} \rangle=1$ is positive, (\ref{triangle}) holds
for the first steps until one of the quantities in (\ref{triangle})
becomes negative. The smaller the initial angle, the larger the
maximum number of steps allowed before (\ref{triangle}) is
violated. The final state has an overlap $f^K=|\langle
\Psi_1^K|\Psi_2^K \rangle|$, from which the probability of error in
distinguishing $\ket{\Psi_1^K}$ and $\ket{\Psi_2^K}$ can be computed,
using $P_E={1 \over 2}(1-\sqrt{1-(f^K)^2})$ \cite{Hels}. The
measurement that attains this value of $P_E$ is a von Neumann
measurement. In order to carry out the measurement one first makes a
final unitary transformation and then measures which path the
photon takes.

Our strategy is to repeat the protocol until it succeeds (no
absorption occurs). Hence the expected number of photons absorbed for
object $i$ will be \begin{eqnarray}
\label{nabs} \bar N^{abs}_i&=&\sum_{i=1}^\infty
(1-P(abs|i))P(abs|i)^n\nonumber \\
&=&P(abs|i)/(1-P(abs|i)). \nonumber
\end{eqnarray}
where $P(abs|i)$ is the probability of absorbing a photon in the protocol.

\begin{figure}[h]
\centerline {\psfig{width=9.0cm,file=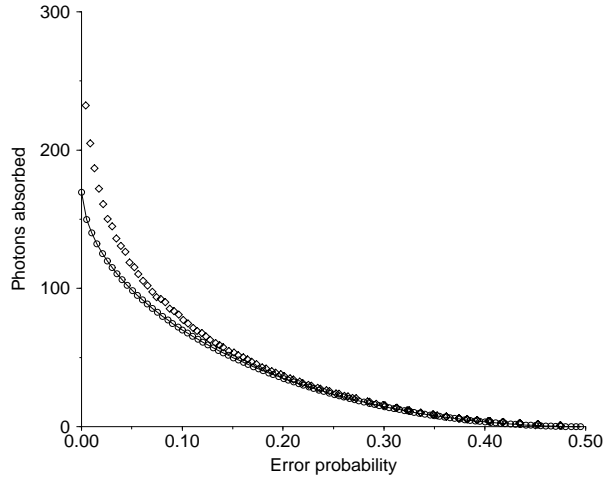}}
\begin{narrowtext}
\caption{ The average number of photons absorbed, $\bar N^{abs}$, for
a given error probability $P_E$, for $\alpha_1=0.2$,
$\alpha_2=0.3$. Each circle represents a protocol given by
(\ref{variable}), with a particular random choice of initial
angle. Protocols were selected by the condition that $\bar N^{abs}$
differed by less than $10^{-4}$ from the bound of (\ref{result}). The
diamonds represent numbers of photons absorbed with the photon
counting protocol described in the text.  Note how the curves diverge
for $P_E \to 0$, since a quantum protocol can distinguish with
certainty between two transparencies with a finite number of absorbed
photons whereas a classical absorption protocol requires an infinite
number of absorbed photons for perfect discrimination.
\label{fig:example1}}
\end{narrowtext}
\end{figure}

\begin{figure}[h]
\centerline {\psfig{width=9.0cm,file=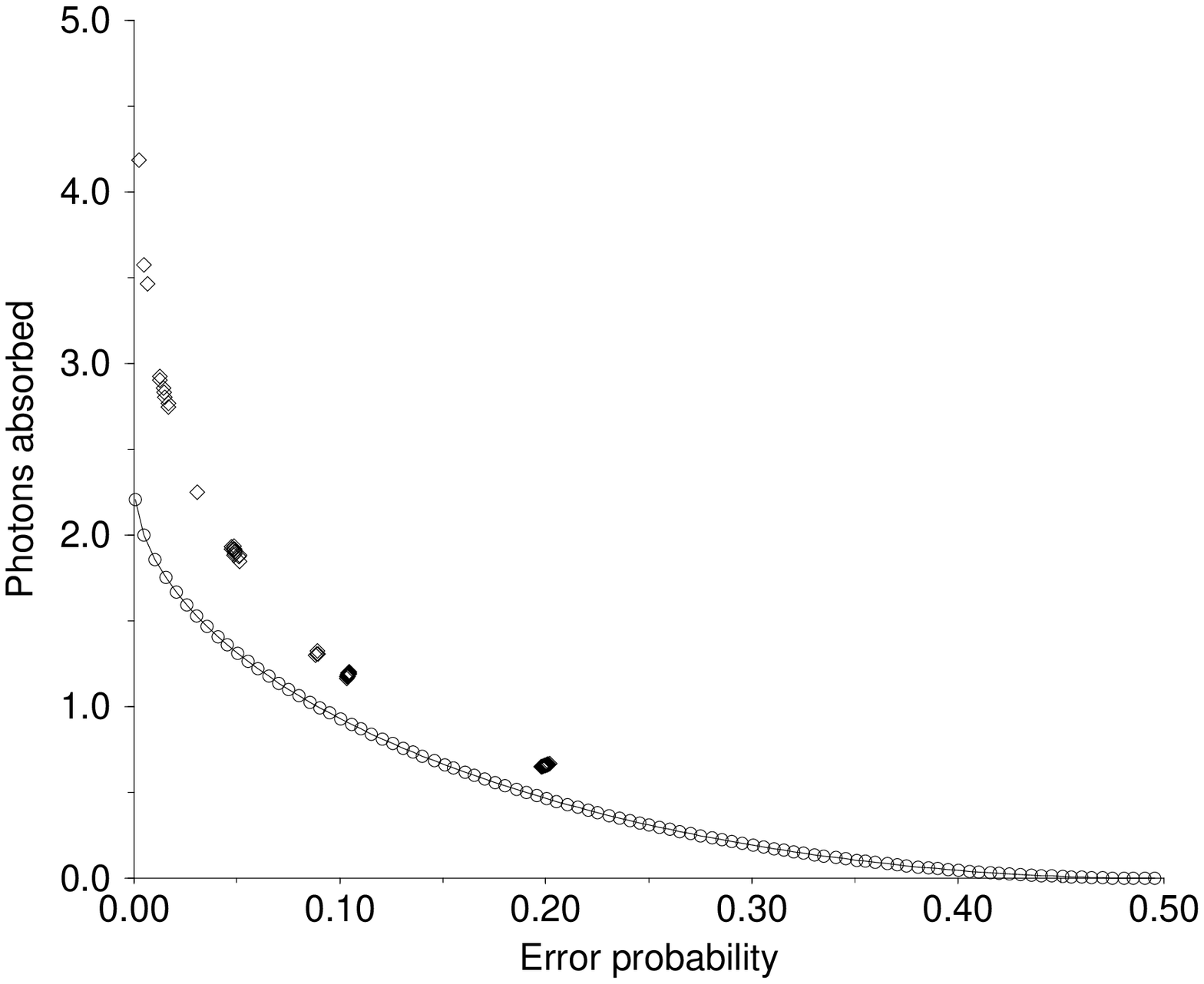}}
\begin{narrowtext}
\caption{ The average number of photons absorbed, as a function of
$P_E$, for $\alpha_1=0.2$, $\alpha_2=0.8$. Notation as in Figure 1.
Note that for the photon counting protocol described in the text the
error probability $P_E$ can only take discrete values. For instance
for the values of $\alpha_{1,2}$ chosen here, if no photons are sent
through the object, $P_E=0.5$ and if a single photon is sent through
the object $P_E=0.2$. Smaller values of $P_E$ require more photons.
\label{fig:example2}}
\end{narrowtext}
\end{figure}

By varying the value of $K$ in the allowed range, i.e. before
(\ref{triangle}) is violated, we get a set of algorithms yielding
certain values of $P_E$. We found that, for all values of $\alpha_1$
and $\alpha_2$ tested, as the initial angle $\theta^0$ was varied, a
complete range of values of $P_E$ from zero to ${1\over 2}$ was
obtained. The small circles in Figures \ref{fig:example1} and
\ref{fig:example2} show this for two values of the $\alpha_i$. As can
be seen, the bound (solid curve) is closely approached for all the
$P_E$.

\begin{figure}[h]
\centerline {\psfig{width=5.0cm,file=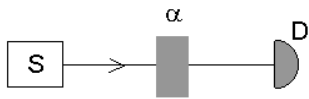}}
\begin{narrowtext}
\caption{A simple classical scheme to distinguish two
transparencies. Photons are sent one by one through the object. The
number of photons that pass through the object is counted. After
each photon is sent, a decision is taken whether it is necessary to
continue to send photons through the object, or whether the error
probability is sufficiently small.
\label{F2}}
\end{narrowtext}
\end{figure}

To compare the quantum bound (\ref{result}) with classical schemes,
figures \ref{fig:example1} and \ref{fig:example2} also show the mean
number of photons absorbed in a photon-counting protocol (illustrated
in figure \ref{F2}). One possible strategy is to send a fixed number
of photons through the specimen and decide, by comparing the number
absorbed with a predetermined threshold, which object is present.  In
general, however, fewer photons need be absorbed if the situation is
appraised after each photon is transmitted. Suppose that, after the
$n$-th photon has been transmitted, $m$ have been absorbed. One
calculates the posterior probability
$P(1|m,n)=\alpha_1^{n-m}\beta_1^m/(\alpha_1^{n-m}\beta_1^m+\alpha_2^{n-m}\beta_
2^m)$, assuming still that $p_1=p_2$, and decides that object 1
is present if $P(1|m,n)>1-x$ and object 2 is present if $P(1|m,n)<x$,
where $x$ is a chosen number between 0 and 1/2; otherwise one
transmits another photon and repeats the procedure. The number $x$ is
therefore the maximum error probability one will tolerate. The actual
probability of error with a given $x$ can be calculated empirically by
averaging over many trials the values of $P(1|m,n)$ when object 2 is
chosen and $1-P(1|m,n)$ when object 1 is chosen. Similarly, the mean
number of photons absorbed is obtained by averaging over many
trials. Varying $x$ then gives the mean number of absorbed photons as
a function of $P_E$. As Figures \ref{fig:example1} and
\ref{fig:example2} show, this photon-counting strategy entails a
greater expected number of absorbed photons than our algorithm,
especially as $P_E$ tends to zero (when the number of photons absorbed
by the counting strategy must tend to infinity). For instance, for
$P_E=0.01$, only $75\%$ of the classical light is needed for
$\alpha_1=0.2$, $\alpha_2=0.3$ and $63\%$ for $\alpha_1=0.2$,
$\alpha_2=0.8$.

In conclusion, for a given probability of error, the mean number of
absorbed photons given by our quantum bound eq. (\ref{result})
is less than that expected
from a simple absorption technique. The bound we have derived is valid
for any quantum protocol, using any number of photons, ancillae,
etc. However, we have shown that, for real transparencies and equal
prior probabilities, a single photon protocol can approch our bound
arbitrary closely. This suggest that using many photons or coherent
light is not as good as using a single photon source. It would be
interesting to know whether similar types of protocols allow our bound
to be saturated in the case of complex transparencies and unequal
prior probabilities. The latter case deserves particular attention,
since the advantage of our bound over the classical limits seems to be
most marked for unequal priors.

\end{multicols}

\begin{thebibliography}{}

\bibitem{EV} Elitzur, A. C. and
Vaidman, L., Found. of Phys.{\bf 23}, 987-997 (1993).



\bibitem{Kwiat} Kwiat, P., Physica Scripta {\bf T76}, 115 (1998).


\bibitem{Jang} Jang, J.-S., Phys. Rev. A, {\bf 59}, 2322-2329 (1999).

\bibitem{KSS} Krenn, G., Summhammer, J. and K. Svozil, Phys. Rev. A {\bf 61},
    052102 (2000)

\bibitem{1} Mitchison, G. and Massar, S.,  Phys, Rev. A {\bf 63} (2001),
032105

\bibitem{MMP} Massar, S., Mitchison, G. and Pironio, S.
preprint available at http://xxx.lanl.gov/quant-ph/0102116.

\bibitem{KW} Kent, A. and Wallace, D.,
preprint available at http://xxx.lanl.gov/quant-ph/0102118.

\bibitem{FHKPR} P. Facchi, Z. Hradil, G. Krenn, S. Pascazio,
  J. \v{R}eh\'{a}\v{c}k,
preprint  available at http://xxx.lanl.gov/quant-ph/0104021.

\bibitem{Zeno} Kwiat, P. G., Weinfurter, H., Herzog, T.,
Zeilinger, A. and Kasevich, M. A.,  Phys. Rev. Lett. {\bf 74},
4763-4766 (1995).


\bibitem{Hels} C. W. Helstrom, {\em Quantum Detection and Estimation
Theory}, (Academic Press, New York,1976)

\end{thebibliography}
\end{document}